\documentstyle[12pt,epsfig,axodraw,a4]{article} 
\newcommand{\vs}{\vspace{-0.25cm}}

\begin{document} 

\begin{center}
\large{\bf Chiral 3\begin{boldmath}$\pi$\end{boldmath}-exchange NN-potentials:
results for representation-invariant classes of diagrams}

\bigskip 

\bigskip

N. Kaiser\\

\bigskip

Physik Department T39, Technische Universit\"{a}t M\"{u}nchen,\\
    D-85747 Garching, Germany

\end{center}

\bigskip

\bigskip

\begin{abstract}
We calculate in (two-loop) chiral perturbation theory the local NN-potentials
generated by certain classes of three-pion exchange diagrams. We
stress that the chiral $3\pi NN$-contact vertex depends on the particular 
choice of the interpolating pion-field and therefore one has to consider 
representation invariant classes of diagrams by supplementing graphs involving
the chiral $4\pi$-vertex. We find that the resulting isovector spin-spin and
tensor NN-potentials are negligibly small for $r\geq 0.8$ fm. One can conclude
that the effects of uncorrelated chiral $3\pi$-exchange in the 
NN-interaction are very small and therefore of no practical relevance.      
\end{abstract}

\bigskip
PACS: 12.20.Ds, 12.38.Bx, 12.39.Fe, 13.75.Cs. 

\bigskip
Accepted for publication in: {\it Physical Review} {\bf C}
\bigskip
\vskip 1.5cm
The longest-range part of the strong nucleon-nucleon interaction is due to the
well-established one-pion exchange. Next in range comes the two-pion exchange
force, whose formulation has been a longstanding problem in quantum
field theory and in dispersion theory. In recent years, it has been argued that
the key to the solution is the chiral symmetry of QCD, and that the long-range
parts of the two-pion exchange NN-potential can be derived model-independently
via a systematic expansion of the effective chiral pion-nucleon Lagrangian. The
complete leading and next-to-leading order terms of this chiral two-pion 
exchange NN-potential have been calculated in ref.\cite{nnpap1} (actually most
of the contributions have already been obtained in earlier calculations).
Recently, the elastic proton-proton scattering data base below 350\,MeV
laboratory kinetic energy (consisting of 1951 data points) has been analyzed in
terms of one-pion exchange and chiral two-pion exchange in ref.\cite{nijmeg}. 
The resulting good $\chi^2/dof \leq 1$ constitutes a convincing proof for the
presence of the chiral two-pion exchange in the long-range proton-proton strong
interaction. Furthermore, it was concluded in ref.\cite{nijmeg} that
$1\pi$-exchange together with chiral $2\pi$-exchange gives a very good NN-force
at least as far inwards as $r=1.4$\,fm inter-nucleon distance.  

Naturally, the next question to be answered is whether there are some 
important effects from chiral three-pion exchange. In a recent work Pupin and 
Robilotta \cite{robil} calculated the potentials generated by one specific 
two-loop diagram in which the three pions are emitted from a contact vertex at
the first nucleon and absorbed on a contact vertex at the second nucleon (first
graph in Fig.\,1). The resulting
isovector spin-spin and tensor potentials were rather small with values of less
than 1.6\,MeV for distances $r\geq 0.8$\,fm, i.e. two to three orders of
magnitude smaller than typical $2\pi$-exchange potentials.  The form of the
chiral $3\pi NN$-contact vertex employed in the calculation of ref.\cite{robil}
stems from an effective chiral Lagrangian using a specific choice of
the interpolating pion-field. As a matter of fact the non-linear realization 
of chiral symmetry \cite{weinb} allows for redefinitions of the pion-field 
and consequently the result of the single diagram studied in ref.\cite{robil}
becomes non-unique. On the level of Feynman diagrams this feature implies that
only suitable classes of graphs, but not individual diagrams can have a
physical meaning. This fact was also stressed in ref.\cite{nnpap2} when
calculating the isoscalar central NN-amplitude generated by correlated chiral
$2\pi$-exchange. The diagram involving the chiral $4\pi$-vertex had to be
supplemented by four additional graphs involving the chiral $3\pi NN$-contact
vertex. The purpose of this work is to present results for the chiral
$3\pi$-exchange NN-potential as given by such complete classes of diagrams. A
calculation of the very large number of all possible (two-loop) $3\pi$-exchange
diagrams goes beyond the scope of the present short paper. In order to learn
about the generic size of uncorrelated $3\pi$-exchange effects in the
NN-interaction we will restrict ourselves here to the evaluation of four
(relatively simple) classes of diagrams. In fact these four classes comprise
all $3\pi$-exchange graphs carrying a common prefactor $g_A^2/f_\pi^6$.   

Let us begin with recalling the effective chiral Lagrangians for $\pi\pi$- 
and $\pi N$-interaction, which read at lowest order     
\begin{equation} {\cal L}_{\pi\pi}^{(2)} = {f_\pi^2 \over 4} {\rm tr} (
\partial^\nu U \partial_\nu U^\dagger+m_\pi^2 (U+U^\dagger))\,,
\end{equation}
\begin{equation} {\cal L}_{\pi N}^{(1)} = \bar N \Big( i D_0-{g_A\over 2}\,\vec
\sigma \cdot \vec u\Big) N\,, \quad u^\nu = i\{ \xi^\dagger, \partial^\nu \xi\}
 \,, \end{equation}
where $f_\pi=92.4$\,MeV is the weak pion decay constant and $g_A=g_{\pi
N}f_\pi/M= 1.32$ as given by the Goldberger-Treiman relation together with 
$g_{\pi N}=13.4$. The choice $g_A=1.32$ is most natural in the present context
since the pion-nucleon vertex is the relevant one here and not the axial-vector
coupling.  ${\cal L}_{\pi N}^{(1)}$ is presented in the heavy
baryon formulation (i.e. a non-relativistic treatment of the nucleons) with 
$\vec \sigma$
denoting the Pauli spin-matrices and $D^\nu = \partial^\nu +{1\over 2}
[\xi^\dagger, \partial^\nu \xi]$ the chiral covariant derivative acting on the
iso-doublet nucleon-field $N$. The $SU(2)$-matrix $U=\xi^2$ collects the
Goldstone pion-fields  in the form 
\begin{equation} U(\vec \pi) = 1+ {i\over  f_\pi } \vec \tau \cdot \vec \pi - 
{1\over 2 f_\pi^2 } {\vec \pi}^2 - {i \alpha \over f_\pi^3 } (\vec \tau \cdot
\vec \pi)^3  + {8\alpha -1 \over 8f_\pi^4 } {\vec \pi}^4 + \dots
\end{equation} 
Note that only the coefficients of the linear and quadratic term in the 
pion-field $\vec \pi$ are fixed by the proper normalization of the kinetic term
and the unitary condition $U^\dagger U=1$. The numerical coefficient $\alpha$
in front of the third power of the pion-field is $arbitrary$ and it reappears 
via the unitarity condition $U^\dagger U=1$ in front of the $\vec \pi^4$-term. 
In fact, when continuing the power series expansion of  $U(\vec \pi)$ in eq.(3)
one  encounters an infinite number of  arbitrary coefficients. This just 
reflects the well-known fact that the non-linear realization of chiral symmetry
\cite{weinb} on the pion-fields is unique only up to an arbitrary function
$f(\vec \pi^2)$. As a consequence of eq.(3) the off-shell $4\pi$-vertex 
\begin{equation} {i\over f_\pi^2} \Big\{ \delta^{ab}\delta^{cd} [ (q_1+q_2)^2
-m_\pi^2 +2\alpha (4m_\pi^2-q_1^2-q_2^2-q_3^2-q_4^2) ] + two\,\, cycl.\,\,
perm.\,\, \Big\} \end{equation} 
and the chiral $3\pi NN$-vertex (i.e. $N\to
N+\pi^a(q_1)+\pi^b(q_2)+\pi^c(q_3)$), 
\begin{equation} {g_A \over 4f_\pi^3}\, \vec \sigma \cdot \Big\{ \tau^a
\delta^{bc} [ 4\alpha \,\vec q_1 +(4\alpha-1)(\vec q_2+\vec q_3\,) ] +two\,\,
cycl.\,\, perm.\,\,\Big\} \end{equation} 
both become $\alpha$-dependent. In (measurable) on-shell matrix elements the
unphysical parameter $\alpha$ must of course drop out. For elastic
$\pi\pi$-scattering this is obvious due to the mass-shell condition
$q_1^2=q_2^2=q_3^2=q_4^2=m_\pi^2$. The T-matrix for the reaction $ \pi N\to
\pi\pi N$ at threshold in the center-of-mass frame    
\begin{equation} {\cal T}_{\rm th}^{\rm cm}(\pi^a(\vec k\,)N\to \pi^b\pi^c N) 
=i\,\vec\sigma \cdot \vec k \, \Big[ D_1(\tau^b \delta^{ac}+\tau^c
\delta^{ab})+ D_2 \,\tau^a \delta^{bc} \Big]\,, \end{equation} 
receives contributions from the chiral $3\pi NN$-contact vertex eq.(5) and the
pion-pole diagram of the form 
\begin{equation}D_1^{\rm cont}= {g_A \over 4 f_\pi^3} (4\alpha-1)\,, \quad 
D_1^{\rm \pi-pole }= {g_A \over 8 f_\pi^3} (3-8\alpha)\,,\end{equation} 
\begin{equation}  \quad D_2^{\rm cont} ={g_A \over  f_\pi^3} \alpha\,,
\quad D_2^{\rm \pi-pole }= -{g_A \over8 f_\pi^3} (8\alpha+3)\,. \end{equation}
The sums $D_{1,2}^{(\rm cont)}+D_{1,2}^{(\rm\pi-pole)}$ are indeed
$\alpha$-independent and they constitute the leading order terms of the chiral 
low-energy theorems for $\pi N\to \pi\pi N$ derived in ref.\cite{ppnlet}. Note
that there is no value of $\alpha$ which would allow one to derive the complete
leading order terms for $D_1$ and $D_2$ from a single diagram. Graphs with the
chiral $3\pi NN$-contact vertex and graphs with the chiral $4\pi$-vertex always
have to be grouped into classes and only the results of such classes of
diagrams have a physical meaning.  

%\bigskip
%\bigskip

%\bild{3pipotfig1.epsi}{16}
%\smallskip
%{\it Fig.1: $3\pi$-exchange diagrams of class\,I. Solid and dashed lines
%represent nucleons and pions, respectively. The symmetry factor of these 
%graphs is 1/6.}

%\bigskip

\vskip 2cm
\begin{center}

\SetWidth{1.5}
  \begin{picture}(400,70)
\Line(5,0)(5,85)
\Line(80,0)(80,85)
\DashLine(5,42.5)(80,42.5){6}
\DashCurve{(5,42.5)(42.5,67.5)(80,42.5)}{6}
\DashCurve{(5,42.5)(42.5,17.5)(80,42.5)}{6}
\Vertex(5,42.5){3}
\Vertex(80,42.5){3}

\Line(115,0)(115,85)
\Line(190,0)(190,85)
\DashLine(115,42.5)(190,42.5){6}
\DashCurve{(145,42.5)(167.5,67.5)(190,42.5)}{6}
\DashCurve{(145,42.5)(167.5,17.5)(190,42.5)}{6}
\Vertex(115,42.5){3}
\Vertex(190,42.5){3}
\Vertex(145,42.5){3}

\Line(225,0)(225,85)
\Line(300,0)(300,85)
\DashLine(225,42.5)(300,42.5){6}
\DashCurve{(225,42.5)(247.5,67.5)(270,42.5)}{6}
\DashCurve{(225,42.5)(247.5,17.5)(270,42.5)}{6}
\Vertex(225,42.5){3}
\Vertex(300,42.5){3}
\Vertex(270,42.5){3}

\Line(335,0)(335,85)
\Line(410,0)(410,85)
\DashLine(335,42.5)(410,42.5){6}
\DashCurve{(355,42.5)(372.5,67.5)(390,42.5)}{6}
\DashCurve{(355,42.5)(372.5,17.5)(390,42.5)}{6}
\Vertex(335,42.5){3}
\Vertex(355,42.5){3}
\Vertex(390,42.5){3}
\Vertex(410,42.5){3}

  \end{picture}
\end{center}
\medskip
\noindent
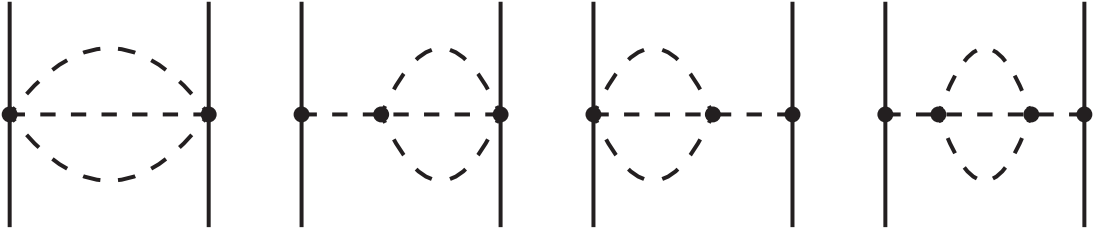
{\it Fig.1: $3\pi$-exchange diagrams of class\,I. Solid and dashed lines
represent nucleons and pions, respectively. The symmetry factor of these graphs
is 1/6.}

\bigskip
\bigskip

Let us now turn to the evaluation of (parts of) the chiral $3\pi$-exchange
NN-potential. According to the previous discussion the full class of two-loop
diagrams shown in Fig.\,1 should be considered as one entity, whereas in 
ref.\cite{robil} only the first one was evaluated for $\alpha=0$. From a 
consideration of the spin- and isospin-factors occurring in these diagrams one
finds immediately that only non-vanishing isovector spin-spin and tensor 
NN-amplitudes  will be obtained, i.e. a contribution to the NN T-matrix of the 
form   
\begin{equation} {\cal T}_{NN} = \Big[ W_S(q)\,\vec \sigma_1\cdot \vec \sigma_2
+ W_T(q) \,\vec \sigma_1\cdot \vec q \,\,  \vec \sigma_2 \cdot \vec q \, \Big] 
\,\vec \tau_1 \cdot \vec \tau_2\,, \end{equation} 
where $q=|\vec q\,|$ denotes the momentum transfer between the initial and
final state nucleon. Obviously, the two-loop pion-pole diagrams in Fig.\,1
contribute via mass and coupling constant renormalization also to the
$1\pi$-exchange. These effects are, however, automatically taken care of by 
working with the physical pion mass $m_\pi$ and physical $\pi NN$-coupling 
constant $g_{\pi N}$. We are interested here only in the coordinate space 
potentials generated by the simultaneous exchange of three pions between both 
nucleons. For that purpose it is sufficient to calculate the imaginary parts of
the NN-amplitudes $W_{S,T}(q)$ analytically continued to time-like momentum 
transfer $q=i\mu-0^+$ with $\mu\geq 3m_\pi$. These imaginary parts are then the
mass-spectra entering  a representation of the local coordinate space 
potentials in form of a continuous superposition of Yukawa  functions,    
\begin{equation} \widetilde W_S(r) = {1\over 6\pi^2 r} \int_{3m_\pi}^\infty
d\mu \,\mu \,e^{-\mu r} \Big[ \mu^2\, {\rm Im}\, W_T(i\mu) - 3\, {\rm Im}\, 
W_S(i\mu) \Big]\,, \end{equation} 
\begin{equation} \widetilde W_T(r) = {1\over 6\pi^2 r^3} \int_{3m_\pi}^\infty
d\mu\,\mu\, e^{-\mu r}(3+3\mu r+\mu^2 r^2){\rm Im}\, W_T(i\mu)\,.\end{equation}
The spin-spin and tensor potentials, denoted here $\widetilde W_{S,T}(r)$, are
as usual that ones which are accompanied by the operators $\vec \sigma_1\cdot
\vec \sigma_2$ and  $3\,\vec \sigma_1\cdot \hat r\,\vec \sigma_2 \cdot \hat r 
-\vec \sigma_1\cdot \vec \sigma_2$, respectively.

Application of the Cutkosky cutting-rules gives the imaginary parts
Im\,$W_{S,T}(i\mu)$ as integrals of the squared $\bar NN\to 3\pi$ transition
amplitudes over the Lorentz-invariant three-pion phase space. Some details 
about these techniques can be found in ref.\cite{bkm} where the method has been
used to calculate the spectral-functions of the isoscalar electromagnetic and
isovector axial form factors of the nucleon. We find the following result from
the diagrams of class\,I shown in Fig.1,    
\begin{equation} {\rm Im}\, W_S^{(I)}(i\mu) = {g_A^2 \over 6\mu^4(16\pi f_\pi^2
)^3 } \int_{2m_\pi}^{\mu-m_\pi} dw \sqrt{(w^2-4m_\pi^2)\lambda^3(w^2,m_\pi^2,
\mu^2) }\,, \end{equation} 
\begin{eqnarray} {\rm Im}\, W_T^{(I)}(i\mu) &=& {2g_A^2(\mu^2-m_\pi^2)^{-2} 
\over 9(16\pi f_\pi^2\mu^2)^3 } \int_{2m_\pi}^{\mu-m_\pi} dw \sqrt{(w^2-4
m_\pi^2) \lambda (w^2,m_\pi^2,\mu^2) } \\ \nonumber & & \times\Big[3w^4  
(7\mu^4+4\mu^2  m_\pi^2 +m_\pi^4) -2\mu^8-19\mu^6 m_\pi^2-13\mu^4m_\pi^4-17
\mu^2m_\pi^6-3m_\pi^8 \Big]\,, \end{eqnarray}
with $w$ the invariant mass of a pion-pair and $\lambda(x,y,z)= x^2+y^2+z^2-
2xy-2xz-2yz$ the conventional K\"allen function. We note that the spin-spin
part Im\,$W_S^{(I)}(i\mu)$ comes solely from the first graph in Fig.1 
independent of the parameter $\alpha$. In order to compare with the calculation
of ref.\cite{robil} (i.e. the first diagram in Fig.1 for $\alpha=0$) one merely
has to replace the polynomial in the square bracket of eq.(13) by
$(\mu^2-m_\pi^2)^2 (3w^4+ 10\mu^4-5\mu^2m_\pi^2-3m_\pi^4)$. In the chiral 
limit, $m_\pi=0$, all integrals can be performed analytically and one obtains
$3\pi$-exchange potentials with a simple $r^{-7}$-dependence,
\begin{equation} \widetilde W_T^{(I)}(r) = 7\, \widetilde W_S^{(I)}(r) = {70 
g_A^2 \over 3(8\pi)^5 f_\pi^6 }\,{1\over r^7}\,. \end{equation}
One observes in the chiral limit that the tensor and spin-spin potentials of
the complete class\,I are a factor 7  and 25 smaller than the ones of
ref.\cite{robil}. This indicates that there are  large cancelations between
individual diagrams in class\,I. The asymptotic fall-off of the coordinate
space potentials for $r\to \infty$ is determined by the behavior of the 
mass-spectra near the three-pion threshold $\mu= 3m_\pi$. For class\,I the
mass-spectra grow like: Im\,$W_S^{(I)}(i\mu)\sim (\mu-3m_\pi)^3$ and  
Im\,$W_T^{(I)}(i\mu)\sim (\mu-3m_\pi)^2$, and thus one obtains the following 
($r\to \infty$)-asymptotics
\begin{equation}\widetilde W_T^{(I)}(r)=\widetilde W_S^{(I)}(r)={5 g_A^2m_\pi^3
\over 3\sqrt{3}(16\pi )^4 f_\pi^6}\, {e^{-3m_\pi r} \over r^4}+ \dots
\end{equation}
Again, these expressions are a factor 64 smaller than the ones of
ref.\cite{robil} because of cancelations between different diagrams.  
Of course, we have recovered all analytical and numerical results of 
ref.\cite{robil} with our method using the corresponding imaginary parts
mentioned above.

\bigskip
\bigskip

%\bild{3pipotfig2.epsi}{16}
%\smallskip
%{\it Fig.2: $3\pi$-exchange diagrams of class\,II. Diagrams for which the 
%role of both nucleons is interchanged are not shown. They lead to the same
%contribution to the NN-potential. The symmetry factor of these graphs is 1/2.}

%\bigskip

\begin{center}

\SetWidth{1.5}
  \begin{picture}(400,70)
\Line(5,0)(5,85)
\Line(80,0)(80,85)
\DashLine(5,42.5)(80,42.5){6}
\DashCurve{(5,42.5)(42.5,67.5)(80,42.5)}{6}
\DashLine(5,42.5)(80,12.5){6}
\Vertex(5,42.5){3}
\Vertex(80,42.5){3}
\Vertex(80,12.5){3}

\Line(115,0)(115,85)
\Line(190,0)(190,85)
\DashLine(115,42.5)(190,42.5){6}
\DashCurve{(115,42.5)(152.5,17.5)(190,42.5)}{6}
\DashLine(115,42.5)(190,72.5){6}
\Vertex(115,42.5){3}
\Vertex(190,42.5){3}
\Vertex(190,72.5){3}

\Line(225,0)(225,85)
\Line(300,0)(300,85)
\DashLine(225,42.5)(300,42.5){6}
\DashCurve{(255,42.5)(277.5,67.5)(300,42.5)}{6}
\DashLine(255,42.5)(300,12.5){6}
\Vertex(225,42.5){3}
\Vertex(255,42.5){3}
\Vertex(300,42.5){3}
\Vertex(300,12.5){3}

\Line(335,0)(335,85)
\Line(410,0)(410,85)
\DashLine(335,42.5)(410,42.5){6}
\DashCurve{(365,42.5)(387.5,17.5)(410,42.5)}{6}
\DashLine(365,42.5)(410,67.5){6}
\Vertex(335,42.5){3}
\Vertex(365,42.5){3}
\Vertex(410,42.5){3}
\Vertex(410,67.5){3}

  \end{picture}
\end{center}
\medskip
\noindent
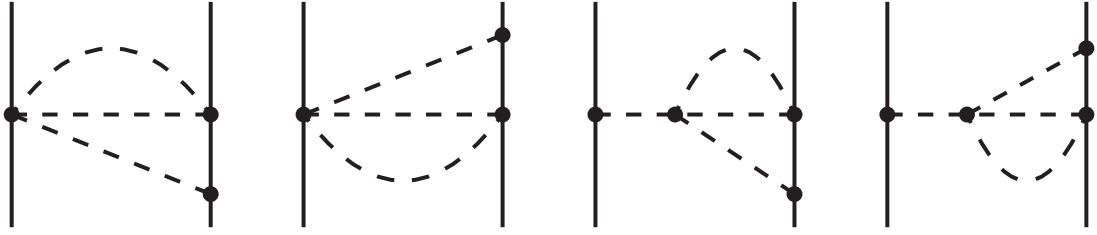
{\it Fig.2: $3\pi$-exchange diagrams of class\,II. Diagrams for which the role 
of both nucleons is interchanged are not shown. They lead to the same
contribution to the NN-potential. The symmetry factor of these graphs is 1/2.}

\bigskip
\bigskip

Next, we consider the diagrams of class\,II shown in Fig.\,2. These are the
diagrams with exactly one nucleon-propagator and because of this property the
invariant $3\pi$-phase space integral can still be reduced to a simple
one-dimensional integral in the heavy nucleon mass limit $M\to \infty$ (compare
also with Im\,$G_A(t)$ in ref.\cite{bkm}). After a somewhat lengthy calculation
we find from class\,II,
\begin{equation} {\rm Im}\, W_S^{(II)}(i\mu) = {g_A^2\over 3 \mu^4(16\pi 
f_\pi^2)^3 }  \int_{2m_\pi}^{\mu-m_\pi} dw \sqrt{(w^2-4m_\pi^2)
\lambda (w^2,m_\pi^2,\mu^2) } \, w^2(3m_\pi^2 +\mu^2-3w^2)\,, \end{equation}
\begin{eqnarray} {\rm Im}\, W_T^{(II)}(i\mu) &=& {g_A^2 \over 6(8\pi f_\pi^2
\mu^2)^3 } \int_{2m_\pi}^{\mu-m_\pi} dw \sqrt{(w^2-4m_\pi^2) \lambda(w^2,
m_\pi^2,\mu^2) } \nonumber \\ & & \times \Big[w^4 m_\pi^2(\mu^2-m_\pi^2)^{-1} 
-2w^2\mu^2+ m_\pi^2(\mu^2+m_\pi^2)(3\mu^2+m_\pi^2)w^{-2}\Big]\,.\end{eqnarray}
In the chiral limit, $m_\pi=0$, one obtains now attractive isovector tensor and
spin-spin potentials with a $r^{-7}$-dependence,
\begin{equation} \widetilde W_T^{(II)}(r) = {28\over 13}\,\widetilde W_S^{(II)}
(r) = -{35 g_A^2 \over 18(4\pi)^5 f_\pi^6 }{1\over r^7} \,.\end{equation} 
The asymptotic fall-off for $r\to \infty$ differs from class\,I due to a
different threshold behavior of the mass-spectra: Im\,$W_S^{(II)}(i\mu)\sim
(\mu-3m_\pi)^4$ and Im\,$W_T^{(II)}(i\mu)\sim(\mu-3m_\pi)^3$, which leads to  
\begin{equation}\widetilde W_T^{(II)}(r)=\widetilde W_S^{(II)}(r)=-{14 g_A^2
m_\pi^2 \over 3\sqrt{3}(8\pi )^4 f_\pi^6} {e^{-3m_\pi r} \over r^5}+ \dots
\end{equation}

\bigskip\bigskip

%\bild{3pipotfig3.epsi}{16}
%\smallskip
%{\it Fig.3: $3\pi$-exchange diagrams of class\,III. The isoscalar
%NN-amplitudes sum up to zero.}

%\bigskip

\begin{center}

\SetWidth{1.5}
  \begin{picture}(400,70)
\Line(5,0)(5,85)
\Line(80,0)(80,85)
\DashLine(5,20)(80,20){6}
\DashLine(5,20)(80,65){6}
\DashLine(5,65)(80,65){6}
\Vertex(5,20){3}
\Vertex(80,65){3}
\Vertex(80,20){3}
\Vertex(5,65){3}

\Line(115,0)(115,85)
\Line(190,0)(190,85)
\DashLine(115,20)(190,20){6}
\DashLine(115,65)(190,20){6}
\DashLine(115,65)(190,65){6}
\Vertex(115,20){3}
\Vertex(190,20){3}
\Vertex(190,65){3}
\Vertex(115,65){3}

\Line(225,0)(225,85)
\Line(300,0)(300,85)
\DashLine(225,20)(300,20){6}
\DashLine(225,20)(300,65){6}
\DashLine(225,65)(300,20){6}
\Vertex(225,20){3}
\Vertex(225,65){3}
\Vertex(300,65){3}
\Vertex(300,20){3}

\Line(335,0)(335,85)
\Line(410,0)(410,85)
\DashLine(335,65)(410,65){6}
\DashLine(335,65)(410,20){6}
\DashLine(335,20)(410,65){6}
\Vertex(335,65){3}
\Vertex(335,20){3}
\Vertex(410,65){3}
\Vertex(410,20){3}

  \end{picture}
\end{center}
\medskip
\noindent
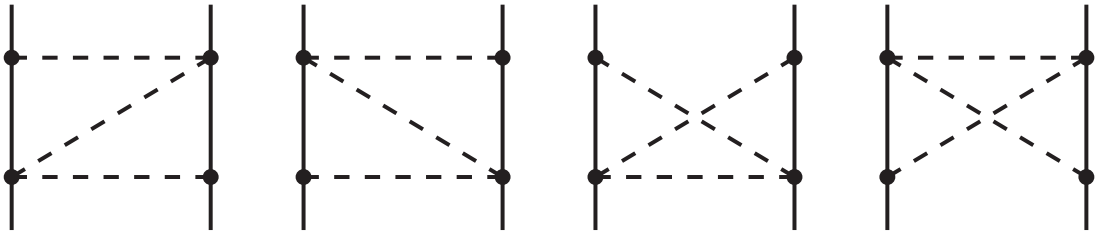
{\it Fig.3: $3\pi$-exchange diagrams of class\,III. The isoscalar NN-amplitudes
sum up to zero.}

\bigskip
\bigskip

Furthermore, we consider the diagrams of class\,III shown in Fig.\,3. The
isospin-factor of the first and second graph is $6-2\,\vec \tau_1 \cdot
\vec\tau_2$ while that of the third and fourth graph is $-6-2\,\vec \tau_1\cdot
\vec\tau_2$. Since all other factors occuring in these two types of diagrams
are equal (modulo the sign of an $i0^+$-term in one heavy nucleon-propagator 
which finally does not matter) one obtains only a contribution to the isovector
spin-spin and tensor NN-amplitudes. Altogether, we find the following imaginary
parts from the diagrams of class\,III,    
\begin{eqnarray} {\rm Im}\, W_S^{(III)}(i\mu) &=& {g_A^2\over 18 \mu^4(16\pi 
f_\pi^2)^3 }  \int_{2m_\pi}^{\mu-m_\pi} dw \sqrt{(w^2-4m_\pi^2)
\lambda (w^2,m_\pi^2,\mu^2) } \nonumber \\ && \times \Big[9w^4-5\mu^4+30 \mu^2
m_\pi^2 -9 m_\pi^4 \Big]\,, \end{eqnarray}
\begin{eqnarray} {\rm Im}\, W_T^{(III)}(i\mu) &=& {2g_A^2 \over 9(16\pi f_\pi^2
\mu^2)^3 } \int_{2m_\pi}^{\mu-m_\pi} dw \sqrt{(w^2-4m_\pi^2) \lambda(w^2,
m_\pi^2,\mu^2) } \nonumber \\ & & \times \Big[4\mu^4 +9\mu^2 m_\pi^2 +9m_\pi^4
 -3w^4 + 6m_\pi^2(\mu^4-m_\pi^4)w^{-2}\Big]\,.\end{eqnarray}
In the chiral limit, $m_\pi=0$, one has again potentials with a
$r^{-7}$-dependence,   
\begin{equation} \widetilde W_T^{(III)}(r) = \widetilde W_S^{(III)}
(r) = {490 g_A^2 \over 9(8\pi)^5 f_\pi^6 }{1\over r^7} \,,\end{equation}
and the asymptotic fall-off of these potentials for $r\to \infty$ is given by 
\begin{equation}\widetilde W_T^{(III)}(r)=\widetilde W_S^{(III)}(r)={g_A^2
m_\pi^3 \over 4\sqrt{3}(4\pi )^4 f_\pi^6} {e^{-3m_\pi r} \over r^4}+ \dots
\end{equation}
\bigskip\bigskip

%\bild{3pipotfig4.epsi}{16}
%\smallskip
%{\it Fig.4: $3\pi$-exchange diagrams of class\,IV. The symmetry factor of 
%these graphs is 1/2.} 

%\bigskip

\begin{center}

\SetWidth{1.5}
  \begin{picture}(400,70)
\Line(5,0)(5,85)
\Line(80,0)(80,85)
\DashLine(5,65)(80,65){6}
\DashCurve{(5,30)(42.5,15)(80,30)}{6}
\DashCurve{(5,30)(42.5,45)(80,30)}{6}
\Vertex(5,30){3}
\Vertex(80,65){3}
\Vertex(80,30){3}
\Vertex(5,65){3}

\Line(115,0)(115,85)
\Line(190,0)(190,85)
\DashLine(115,20)(190,20){6}
\DashCurve{(115,55)(152.5,70)(190,55)}{6}
\DashCurve{(115,55)(152.6,40)(190,55)}{6}
\Vertex(115,20){3}
\Vertex(190,20){3}
\Vertex(190,55){3}
\Vertex(115,55){3}

\Line(225,0)(225,85)
\Line(300,0)(300,85)
\DashCurve{(225,20)(262.5,57.5)(300,65)}{6}
\DashLine(225,65)(300,20){6}
\DashCurve{(225,20)(262.5,27.5)(300,65)}{6}
\Vertex(225,20){3}
\Vertex(225,65){3}
\Vertex(300,65){3}
\Vertex(300,20){3}

\Line(335,0)(335,85)
\Line(410,0)(410,85)
\DashCurve{(335,65)(372.5,57.5)(410,20)}{6}
\DashCurve{(335,65)(372.5,27.5)(410,20)}{6}
\DashLine(335,20)(410,65){6}
\Vertex(335,65){3}
\Vertex(335,20){3}
\Vertex(410,65){3}
\Vertex(410,20){3}

  \end{picture}
\end{center}
\medskip
\noindent
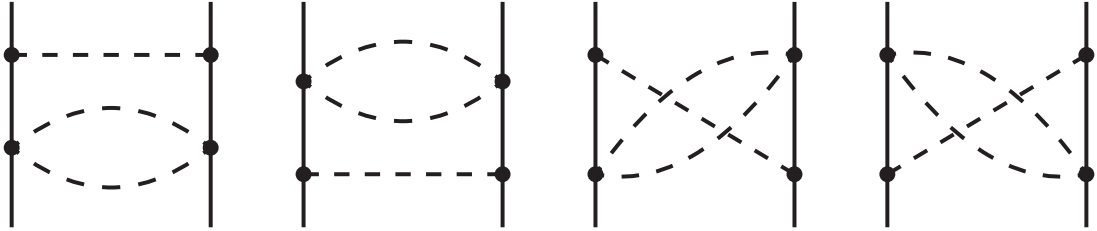
{\it Fig.4: $3\pi$-exchange diagrams of class\,IV. The symmetry factor of these
graphs is 1/2.} 

\bigskip\bigskip

Finally, we consider the diagrams of class\,IV shown in Fig.\,4. The
isospin-factor of the first and second graph (planar boxes) is $6-4\,\vec\tau_1
\cdot \vec\tau_2$ while that of the third and fourth graph (crossed boxes) is
$6 +4\,\vec \tau_1\cdot \vec\tau_2$. In ref.\cite{nnpap1} it was shown that the
irreducible part of the planar box and the crossed box are exactly equal up to
a minus-sign. If one makes here use of this fact, one obtains again only a
contribution to the isovector spin-spin and tensor NN-amplitudes from the
diagrams of class\,IV. The explicit calculation of the corresponding imaginary
parts leads to the following result,     
\begin{eqnarray} {\rm Im}\, W_S^{(IV)}(i\mu) &=& {g_A^2\over 9 \mu^4(16\pi 
f_\pi^2)^3 }  \int_{2m_\pi}^{\mu-m_\pi} dw \sqrt{(w^2-4m_\pi^2)
\lambda (w^2,m_\pi^2,\mu^2) } \nonumber \\ && \times \Big[9w^4-5\mu^4+30 \mu^2
m_\pi^2 -9 m_\pi^4 \Big]\,, \end{eqnarray}
\begin{eqnarray} {\rm Im}\, W_T^{(IV)}(i\mu) &=& {g_A^2 \over 6(8\pi f_\pi^2
\mu^2)^3 } \int_{2m_\pi}^{\mu-m_\pi} dw \sqrt{(w^2-4m_\pi^2) \lambda(w^2,
m_\pi^2,\mu^2) } \,\Big[w^4 +7\mu^2 m_\pi^2  \\ & & -{2\over 3} \mu^4 
-m_\pi^2(\mu^2+m_\pi^2)^2w^{-2}+4m_\pi^2\mu^4(4m_\pi^2-w^2)
\lambda^{-1}(w^2,m_\pi^2,\mu^2) \Big]\,.\nonumber \end{eqnarray}
For the sake of completeness we give the corresponding NN-potentials in the 
chiral limit,  
\begin{equation} \widetilde W_T^{(IV)}(r) = -{7\over 3}\, \widetilde W_S^{(IV)}
(r) = -{140 g_A^2 \over 3(8\pi)^5 f_\pi^6 }{1\over r^7} \,,\end{equation} 
as well as their asymptotic behavior for $r\to \infty$, which reads
\begin{equation}\widetilde W_T^{(IV)}(r)=\widetilde W_S^{(IV)}(r)=-{g_A^2
m_\pi^3 \over 4\sqrt{3}(4\pi )^4 f_\pi^6} {e^{-3m_\pi r} \over r^4}+ \dots
\end{equation}
The remaining chiral $3\pi$-exchange diagrams (not considered here) carry
an independent prefactor $g_A^4/f_\pi^6$ or $g_A^6/f_\pi^6$. Since a larger 
number of nucleon-propagators is involved in these graphs one will be able to 
reduce the relevant $3\pi$-phase space integrals only to double-integrals
of the form $\int \hspace{-1.5mm}\int_{z^2<1} d\omega_1d\omega_2\dots$ (see the
calculation of Im\,$G_E^S(t)$ and Im\,$G_A(t)$ in ref.\cite{bkm}). We hope to
report on these remaining chiral $3\pi$-exchange diagrams scaling with $g_A^4$ 
and $g_A^6$ in a future publication. 
 
In Table\,1, we present numerical results for the isovector spin-spin and
tensor potentials generated by the $3\pi$-exchange graphs of 
class\,I,\,II,\,III and IV for inter-nucleon distances 0.6\,fm\,$\leq
r\leq$\,1.4\,fm, using the parameters
$g_A=1.32$,  $f_\pi=92.4$\,MeV and $m_\pi=138$\,MeV. If one would use instead 
the empirical nucleon axial-vector coupling constant $g_A=1.267\pm0.004$ 
\cite{pdg} the
numerical values would decrease by about 8\%. One observes that the
repulsive isovector spin-spin and tensor potentials coming from the complete
class\,I are much  smaller than the corresponding ones found in
ref.\cite{robil} due to cancelations between different diagrams of class\,I. 
The repulsive isovector  spin-spin and tensor potentials  due to class\,I are
actually overcompensated  by more attractive ones coming from
class\,II. Interestingly, for class\,III the isovector spin-spin and tensor
potential are equal, i.e. $\widetilde W^{(III)}_S(r)=\widetilde
W^{(III)}_T(r)$. This relation is not obvious from the spectral representations
eqs.(10,11) and the imaginary parts given in eqs.(20,21). In all other cases 
one finds that the tensor potential is larger in magnitude than the 
spin-spin  potential. Note that the tensor potentials due to class\,III and
class\,IV cancel each other almost completely. The absolute strengths of the 
$3\pi$-exchange potentials calculated here are less than 0.6\,MeV for distances
$r\geq 0.8$\,fm followed by a very fast exponential fall-off and thus they are
completely negligible in comparison to the $2\pi$-exchange potentials (see
e.g. refs.\cite{nnpap1,nnpap2}). One also notices that the numerical values of
the $3\pi$-exchange potentials at $r=1.4$\,fm deviate still strongly from that
ones given  by the ($r\to \infty$)-asymptotics, eqs.(15,19,23,27). In
particular the spin-spin and tensor potentials due to classes I,\,II,\,IV are 
far from being approximately equal. Similar features were also found in 
ref.\cite{robil} (see Fig.\,5) and there the ($r\to \infty$)-asymptotics turned
out to be accurate on the 20\% level only for $r\geq 4$\,fm. The
($r\to\infty$)-asymptotics is therefore of no use to get a reasonable numerical
estimate of the chiral $3\pi$-exchange potentials for $r\leq 1.5$\,fm.  
 
Furthermore, because of the finite size of the
nucleon one can probably trust a calculation based on point-like nucleons and 
pions only for distances $r\geq 1$\,fm. Moreover, one can expect that the other
chiral $3\pi$-exchange  diagrams not calculated here will produce NN-potentials
larger by at most a factor $g_A^4 \simeq 3$. Such potentials would be still 
much too small in order to be of any practical relevance. One may therefore 
conclude that the effects  due to  uncorrelated (chiral) $3\pi$-exchange in the
NN-interaction are all negligibly small. 
The resonant $3\pi$-exchanges in form
of $\omega(782)$- and $a_1(1260)$-exchange seem to be the dominant effects. For
example, the static  isovector spin-spin and tensor potentials due to 
$a_1(1260)$-exchange, 
\begin{equation} \widetilde W_S^{(a_1)}(r) = -{g_{a_1 N}^2\over 6\pi r}
\,e^{-m_a r}\,, \qquad   \widetilde W_T^{(a_1)}(r) = {g_{a_1 N}^2\over 12\pi r}
\,e^{-m_a r} \Big( 1+{3\over m_a r}+{3\over m_a^2 r^2}\Big)\,, \end{equation} 
are about one order of magnitude larger than  typical chiral $3\pi$-exchange
potentials (see Table\,1). The same holds for the total sum of the isovector
spin-spin and tensor potentials due to the four classes of diagrams evaluated
here. In order to obtain the values of $\widetilde W^{(a_1)}_{S,T}(r)$
given in Table\,1 we used the central value of the $a_1N$-coupling constant 
$g_{a_1 N}^2/4\pi = 7.3\pm 3.0$ as derived from forward
NN-dispersion relations in ref.\cite{kroll}.  The extreme smallness of the 
chiral $3\pi$-exchange NN-potential found here is in agreement
with the results of ref.\cite{bkm}. There it was shown that the
$3\pi$-continuum makes only a negligibly small contribution to the
spectral-functions of the isoscalar electromagnetic and isovector axial form
factors of the nucleon and a few vector-meson poles give therefore a
sufficiently accurate representation of these spectral-functions. As it could
be expected the  same features apply to the NN-potential.

\begin{table}[hbt]
\begin{center}
\begin{tabular}{|c|ccccccccc|}
    \hline

    $r$~[fm]&0.6&0.7&0.8&0.9&1.0&1.1&1.2&1.3 &1.4\\
   \hline
$\widetilde W_S^{(I)}$~[keV]&146 & 38.6& 11.5 & 3.69&1.24&0.43&0.14&0.04&0.01\\
    
$\widetilde W_T^{(I)}$~[keV]& 1591& 476& 164 &62.6&26.0&11.6&5.43&2.67& 1.37\\
\hline    
$\widetilde W_S^{(II)}$~[keV]&--2372 &--736& --262
&--104&--44.5&--20.4&--9.88&--5.00&--2.63\\ 
    
$\widetilde W_T^{(II)}$~[keV]&--4616&--1405&--491 &--191&--80.5&--36.3&
--17.3&--8.63& --4.47\\ \hline
$\widetilde W_S^{(III)}$~[keV]&5121 & 1631& 596 & 242&107&50.1&24.8&12.9&6.93\\
    
$\widetilde W_T^{(III)}$~[keV]& 5121& 1631& 596 &242&107&50.1&24.8&12.9 &6.93\\
    \hline
$\widetilde W_S^{(IV)}$~[keV]&1295 &372& 121 &43.0&16.3&6.49&2.64&1.07 &0.43\\ 
    
$\widetilde W_T^{(IV)}$~[keV]&--4729&--1518&--559 &--228&--101&--47.6&
--23.7&--12.3&--6.63 \\

\hline 
$\widetilde W_S^{(tot)}$~[keV] & 4190& 1306& 467& 185& 80.0& 36.6& 17.7& 9.01&
4.74 \\ 
$\widetilde W_S^{(tot)}$~[keV] &--2633& --816& --290& --114& --48.5& --22.2&
--10.8 & --5.36& --2.80 \\
\hline
$\widetilde W_S^{(a_1)}$~[MeV]&--37.71 &--15.71& --7.26
&--3.41&--1.62&--0.78&--0.38&--0.18 &--0.09\\ 
    
$\widetilde W_T^{(a_1)}$~[MeV]&34.49&14.31& 6.18 & 2.75& 1.25&0.58&
0.27&0.13 &0.06 \\

   \hline
  \end{tabular}
\end{center}
%\medskip 
{\it Tab.1: Isovector spin-spin and tensor NN-potentials due to the
chiral $3\pi$-exchange graphs of classes\,I,\,II,\,III,\,IV (shown in
Figs.\,1,\,2,\,3,\,4) versus the nucleon distance $r$. Note that the
units for these potentials are keV. The total sums of potentials due to
these four classes are denoted by $\widetilde W_{S,T}^{(tot)}$. The 
static $a_1(1260)$-exchange spin-spin and tensor NN-potentials are instead
given in units of MeV.}  
\end{table}

\end{document}